\newcommand{\Tr}{\mathop{\mathrm{Tr}}\nolimits}
\newcommand{\openone}{\leavevmode\hbox{\small1\normalsize\kern-.33em1}}
\newcommand{\bra}[1]{ \langle #1 |}
\newcommand{\ket}[1]{|  #1 \rangle}
\newcommand{\pgen}{\mathbb{P}}
\newcommand{\phs}{\mathbb{P}_{\rm HS}}
\newcommand{\pb}{\mathbb{P}_{\rm B}}
\newcommand{\pc}{\mathbb{P}_{\rm C}}
\newcommand{\phsb}{\mathbb{P}_{\rm HSb}}
\newcommand{\pbb}{\mathbb{P}_{\rm Bb}}
\newcommand{\pcb}{\mathbb{P}_{\rm Cb}}
\newcommand{\pq}{\mathbb{P}_{Q}}
\newcommand{\pd}{\mathbb{P}_{\rm d}}
\newcommand{\pp}{\mathbb{P}_{\rm p}}
\newcommand{\szero}{\hat{S}_0}
\newcommand{\sx}{\hat{S}_x}
\newcommand{\sy}{\hat{S}_y}
\newcommand{\sz}{\hat{S}_z}

\newcommand{\rhounpol}{\hat{\sigma}}
\newcommand{\density}{\hat{\varrho}}
\newcommand{\tr}{{\rm Tr}}
\newcommand{\upol}{\hat{U}_{{\rm pol}}}

\def\proj#1#2{\ket{#1}\bra{#2}}
\newcommand{\eigen}{\lambda}

\documentclass[final,5p,times]{elsarticle}
\usepackage{graphicx,amssymb,bm,subeqnarray,color,ulem}

\journal{Optics Communications}

\begin{document}

\begin{frontmatter}

\title{Quantum degrees of polarization}

\author[1]{Gunnar~Bj\"{o}rk}
\author[1,2]{Jonas~S\"{o}derholm}
\author[2]{Luis~L.~S\'{a}nchez-Soto}
\author[3]{Andrei~B.~Klimov}
\author[4]{Iulia~Ghiu}
\author[4]{Paulina~Marian}
\author[4]{Tudor~A.~Marian}

\address[1]{School of Communication and Information Technology,
Royal Institute of Technology (KTH),
Electrum 229, SE-164 40 Kista, Sweden}

\address[2]{Departamento de \'{O}ptica,
Facultad de F\'{\i}sica,
Universidad Complutense, 28040~Madrid, Spain}

\address[3]{Departamento de F\'{\i}sica,
Universidad de Guadalajara,
44420~Guadalajara, Jalisco, Mexico}

\address[4]{Centre for Advanced Quantum Physics,
University of Bucharest, PO Box MG-11, R-077125,
Bucharest-M\u{a}gurele, Romania}

\begin{abstract}
  We discuss different proposals for the degree of polarization of
  quantum fields. The simplest approach, namely making a
  direct analogy with the classical description via the Stokes
  operators, is known to produce unsatisfactory results. Still, we argue that
  these operators and their properties should be basic for any measure
  of polarization. We compare alternative quantum degrees and put
  forth that they order various states differently. This is to be
  expected, since, despite being rooted in the Stokes
  operators, each of these measures only captures certain characteristics. Therefore, it is
  likely that several quantum degrees of polarization will coexist,
  each one having its specific domain of usefulness.
\end{abstract}

\begin{keyword}
Polarization, Quantum optics, Stokes operators,
Degree of polarization, Unpolarized states

\PACS 42.50.Dv \sep 03.65.Yz \sep 03.65.Ca \sep 42.25.Ja
\end{keyword}

\end{frontmatter}

%% \linenumbers

\section{Introduction}
\label{Sec:Introduction}

Far from a source, all propagating electromagnetic fields can be
treated, to a very good approximation, like plane waves. As a
consequence, for a monochromatic component at a fixed space point,
the tip of the electric field vector describes an ellipse in the
plane transverse to the propagating direction. This geometric
observation has led to the concept of light polarization, which was
laid down already in 1852 in the seminal work of
Stokes~\cite{Sto52}.

Polarization is important in a  variety of classical optical
phenomena, which are used in many applications including remote
sensing~\cite{schott}, light scattering~\cite{barron}, thin-film
ellipsometry~\cite{azzam}, and near-field microscopy~\cite{werner},
to cite only some relevant examples.

The description of polarization for quantum fields has also
attracted a great deal of attention in recent
years~\cite{klyshko,klyshko1997,alojants,luis2000,klimov,luis
review}, mainly due to the rapid growth of quantum information
science. Light is an excellent information carrier as the coded
information remains relatively intact upon propagation, since
photons are very resilient against (unwanted) interactions with the
environment. For example, for visible light at room temperature, the
ratio $\hbar \omega/k_B T$ is approximately equal to 80, so thermal
noise is negligible.  In addition, in an optical fiber, the
absorption is only about 50\% per 10 km of propagation distance at
wavelengths around 1.55~$\mu$m.

Since photon polarization is a property that can be accurately,
rapidly, and almost losslessly manipulated, it is the variable of
choice in many experiments and demonstrations in quantum optics.
Examples include quantum key distribution~\cite{Charly,muller}, quantum
dense coding~\cite{mattle}, polarization entanglement~\cite{kwiat},
quantum teleportation~\cite{bouwmeester}, quantum
tomography~\cite{francesco}, rotationally invariant
states~\cite{radmark}, and phase super-resolution~\cite{resch}.  All
this seems to call for a proper description and quantification of
polarization for quantum fields: our aim here is to make, at
least, a rudimentary overview of such recent developments.

The paper is organized as follows. In Section~\ref{Sec:Stokes}, we
describe the simplest ``translation'' from a classical to a quantum
description in terms of Stokes operators. We point out the problems
arising in this approach, mainly due to the fact that a large set of
states are classified as unpolarized, although they carry some
polarization information. In consequence, we delineate the
conditions for the appearance of this ``hidden'' polarization. In
Section~\ref{Sec:Desiderata}, we discuss criteria and desiderata for
any quantum measure of polarization.  We apply them in
Section~\ref{Sec:Examples} to some distance-based measures, examining
how they may be modified to avoid potential shortcomings. In terms
of these new measures, we investigate the degree of polarization for
maximally polarized pure states. For completeness, we also treat
other non-distance-based degrees of polarization. In Section~\ref{Sec:
nonlinear}, we speculate about how nonlinear transformations would
affect some aspects of this picture. Finally, in
Section~\ref{Sec:discussion conclusions}, we round of the expos\'{e}
with some general remarks and conclusions.

\section{Stokes description of polarization}
\label{Sec:Stokes}

\subsection{Stokes parameters and operators}
\label{Sec:classical}

Let us start by briefly discussing some basic concepts about classical
and quantum polarization. We assume a monochromatic plane wave, whose
electric field lies in the plane perpendicular to its direction of
propagation. Under these conditions, the field can be represented by
two complex amplitudes denoted by $E_H$ and $E_V$ when using the basis
of linear horizontal and vertical polarizations. The Stokes parameters
are then defined as

\begin{eqnarray}
  \label{Stokpar}
  & S_0 = E^*_H E_H + E^*_V E_V \, ,
  \qquad
  S_x = E_H E^*_V + E^*_H E_V \, , &
  \nonumber \\
  & &  \\
  & S_y = i ( E_H E^*_V - E^*_H E_V ) \, ,
  \qquad
  S_z = E^*_H E_H - E^*_V E_V \, . &
  \nonumber
\end{eqnarray}
In case of stochastic fields, one usually uses the average values
given by the corresponding statistical mixture of deterministic
waves. For quantum fields, the amplitudes $E_H$ and $E_V$ are
represented by complex amplitude operators, denoted by $\hat{a}_H$
and $\hat{a}_V$. They obey the bosonic commutation relations
\begin{equation}
  [ \hat{a}_j, \hat{a}_k^\dagger ] = \delta_{jk} \, ,
  \qquad j, k \in \{H, V \} \, .
\end{equation}
The Stokes operators are subsequently introduced as the quantum
counterparts of the classical variables, namely~\cite{collett}
\begin{eqnarray}
  \label{Stokop}
  & \szero = \hat{a}^\dagger_H \hat{a}_H +
  \hat{a}^\dagger_V \hat{a}_V \, ,
  \qquad
  \sx = \hat{a}_H \hat{a}^\dagger_V + \hat{a}^\dagger_H \hat{a}_V \, , &
  \nonumber \\
  & & \\
  & \sy = i ( \hat{a}_H \hat{a}^\dagger_V -
  \hat{a}^\dagger_H \hat{a}_V ) \, ,
  \qquad
  \sz = \hat{a}^\dagger_H \hat{a}_H -
  \hat{a}^\dagger_V \hat{a}_V \, , &
  \nonumber
\end{eqnarray}
and their mean values correspond to the Stokes parameters
($\langle \hat{S}_0 \rangle$, $\langle \hat{\mathbf{S}} \rangle$), where
$\hat{\mathbf{S}} = (\hat{S}_x, \hat{S}_y, \hat{S}_z)$. The Stokes
operators satisfy the SU(2)-like commutation relations:
\begin{equation}
  \label{eq:ccrsu2}
  [ \hat{S}_{x}, \hat{S}_{y}]  = 2 i \hat{S}_z \, ,
\end{equation}
and cyclic permutations. The noncommutability of these operators
precludes the simultaneous exact measurement of the corresponding
physical quantities. Among other consequences, this implies that no
field state (leaving aside the two-mode vacuum) can have definite
nonfluctuating values of all the Stokes operators simultaneously.
This is expressed by the uncertainty relation
\begin{equation}
  (\Delta \mathbf{S})^2  =
  (\Delta S_x)^2  + (\Delta S_y)^2 + (\Delta S_z)^2 \geq 2
  \langle \hat{S}_0 \rangle \, ,
\end{equation}
where the variances are $(\Delta X)^2 = \langle \hat{X}^2 \rangle -
\langle \hat{X} \rangle^2$. This reflects the fact that, contrary to
the classical optics description, the electric field of a
monochromatic field never describes a definite ellipse in its
quantized description.

Using Stokes operators, the standard degree of polarization employed
in classical optics can be generalized to quantum fields through the
definition~\cite{Bro1998}
\begin{equation}
  \mathbb{P}_{\mathrm{S}} =
  \frac{| \langle \hat{\mathbf{S}} \rangle |}
  {\langle \hat{S}_0 \rangle} =
  \frac{\sqrt{\langle \hat{S}_x \rangle^2
   + \langle \hat{S}_y \rangle^2
   + \langle \hat{S}_z \rangle^2}}
  {\langle \hat{S}_0 \rangle} \, .
  \label{Eq:PsDef}
\end{equation}
We will refer to this definition as the Stokes degree of
polarization, since it is the length of the normalized Stokes vector
(so $0 \le \mathbb{P}_{\mathrm{S}} \le 1$). Expression
(\ref{Eq:PsDef}) is undefined for $\langle \hat{S}_{0} \rangle = 0$,
i.e., when both modes are in the vacuum state. However, in order to
simplify our discussion below, we complement definition
(\ref{Eq:PsDef}) with $\mathbb{P}_\mathrm{S} \equiv 0$ for the
two-mode vacuum. We also note in passing that $\mathbb{P}_{\mathrm{S}}$
depends exclusively on the first moments of the Stokes operators.
However, it follows from the relation
\begin{equation}
  \hat{\mathbf{S}}^2 = \szero (\szero + 2)
  \label{Eq:Svec2}
\end{equation}
that $\mathbb{P}_{\mathrm{S}}$ can be recast as \cite{luis review}
\begin{equation}
  \mathbb{P}_{\mathrm{S}} =
  \frac{\sqrt{\langle \hat{S}_0 ( \hat{S}_0 + 2 ) \rangle
      - (\Delta \mathbf{S})^2 }}
  {\langle \hat{S}_0 \rangle} \, ,
\end{equation}
which shows that it can be expressed either in terms of
the average Stokes vector or its fluctuations.

We observe that $\hat{S}_{0} = \hat{N}_{H} + \hat{N}_{V}$, where
$\hat{N}_{H}$ ($\hat{N}_{V}$) is the photon number operator in mode $H$
($V$) and that \begin{equation}
  \label{eq:ccs0}
  [\hat{S}_{0}, \hat{\mathbf{S}} ] = 0 \, ,
\end{equation}
so each energy manifold can be treated separately. To bring out
this point more clearly, it is advantageous to relabel the
standard two-mode Fock basis as
\begin{equation}
  \label{eq:invsub}
  | k, N - k \rangle = |k \rangle_{H} \otimes |N - k \rangle_{V} ,
  \qquad
  k= 0, 1, \ldots, N \, ,
\end{equation}
so that, for each fixed total number of photons $N$, these states span
an SU(2) invariant subspace of dimension $N + 1$.

\subsection{Hidden polarization}
\label{Sec:hidden}

As noticed early on, there are problems with the definition
(\ref{Eq:PsDef}). For example, this approach assigns zero degree of
polarization to pure fields that carry polarization information.
This is referred to as ``hidden
polarization''~\cite{klyshko,klyshko1997}, but perhaps it would be
better to say that such states have higher-order polarization.
Classical fields can also have significant higher-order polarization
correlations. There are, for example, stochastic classical fields
that can be seen as statistical mixtures of fully polarized states,
and simultaneously be unpolarized according to its average Stokes
vector. In order to fully characterize such classical mixtures, one
would need higher-order moments of the Stokes operators.

In the literature it becomes quite clear that the vast majority of
physicists view the classical counterpart of Eq.~(\ref{Eq:PsDef}) as
{\it the} degree of polarization of a plane-wave classical field. In
the quantum physics community it has been common to measure
higher-order moments, and hence, the inadequacies of the definition
(\ref{Eq:PsDef}) has been more visible in this community. However,
$\mathbb{P}_{\mathrm{S}}$ assigns a relevant degree of polarization
to every pure state in classical optics, whereas this is not the
case in quantum optics.

For a state to have $\mathbb{P}_{\rm S} = 0$, the expectation values
of the Stokes vector $\hat{\mathbf{S}}$ must vanish. To derive the
set of pure $N$-photon states that are unpolarized according to the
Stokes definition, let
\begin{equation}
  \ket{\Psi_N} = \sum_{k=0}^N c_k \ket{k,N-k} \, ,
  \qquad
  \sum_{k=0}^N |c_k|^2 = 1 \, .
  \label{Eq:arbitrary N-state}
\end{equation}
denote a general, normalized, pure $N$-photon state.  Since
$\langle \hat{a}_H^\dagger \hat{a}_V \rangle = \langle \hat{a}_H
\hat{a}_V^\dagger \rangle^\ast$, we find that
\begin{equation}
  \bra{\Psi_N} \hat{a}_H^\dagger \hat{a}_V
  \ket{\Psi_N} = \sum_{k=0}^{N-1} c_k c_{k+1}^\ast \sqrt{(k+1)(N-k)}
  = 0
  \label{Eq:UnpolCondS12}
\end{equation}
is a necessary and sufficient condition for $\langle \hat{S}_{x}
\rangle$ and $\langle \hat{S}_{y} \rangle$ to vanish simultaneously.
To achieve $\mathbb{P}_{\rm S} = 0$, we must also have
\begin{equation}
  \bra{\Psi_N} \sz \ket{\Psi_N} = \sum_{k=0}^N |c_k|^2 (2 k - N) = 0 .
  \label{Eq:UnpolCondS3}
\end{equation}
Equations (\ref{Eq:UnpolCondS12})-(\ref{Eq:UnpolCondS3}) are thus
necessary and sufficient conditions for the Stokes degree of
polarization of a pure $N$-photon state to vanish. Clearly,
$N$-photon states that have photon-distribution probabilities with
the horizontal-vertical symmetry $|c_k|^2 = |c_{N-k}|^2$ satisfy
$\langle \hat{S}_z \rangle = 0$. Examples of Stokes unpolarized
states with this symmetry in any odd manifold $N \geq 5$ and any
even manifold are given by states satisfying
\begin{equation}
  c_{N-k} = \pm (-1)^k i c_k^\ast ,
  \label{eq:UnpolarizedSymmetricStates}
\end{equation}
where the upper or lower sign is used for all $k$, and $c_{(N \pm
1)/2} = 0$ for odd $N$. For even $N$, the solutions corresponding to
the upper and lower sign imply $\arg c_{N/2} = (N - 1 \pm 2) \pi/4$
and $\arg c_{N/2} = (N + 1 \pm 2) \pi/4$, respectively.

In excitation manifold $N=0$ there exists only one state, the
two-mode vacuum state $\ket{0,0}$, and it fulfills
Eqs.~(\ref{Eq:UnpolCondS12}) and (\ref{Eq:UnpolCondS3}) and thus has
$\mathbb{P}_{\mathrm{S}}=0$ in accordance with our complement to
definition (\ref{Eq:PsDef}).

All pure single-photon states lie on the surface of the
Poin\-ca\-r\'{e} sphere. That is, the corresponding vectors $\langle
\hat{\mathbf{S}} \rangle$ have unit length and are thus fully
polarized ($\mathbb{P}_{\rm S} = 1$). Hence, no Stokes unpolarized
pure state exists in manifold $N=1$.

Since an overall phase factor has no physical significance,
Eqs.~(\ref{Eq:UnpolCondS12}) and (\ref{Eq:UnpolCondS3}) imply that
any Stokes unpolarized pure state in manifold $N = 2$ can be written
as
\begin{equation}
  a e^{i \theta} \ket{0,2} + i \sqrt{1 - 2 a^2} \ket{1,1} + a e^{-i
    \theta} \ket{2,0} ,
  \label{Eq:2phUnpol}
\end{equation}
where $a$ and $\theta$ are real numbers and $0 \leq a \leq
1/\sqrt{2}$. That is, they are of the form
(\ref{eq:UnpolarizedSymmetricStates}). Although all unpolarized
states have vanishing Stokes parameters according to the definition
of $\mathbb{P}_{\rm S}$, the corresponding fluctuations are, in
general, anisotropic. Explicitly, the state (\ref{Eq:2phUnpol}) has
the variances
\begin{subeqnarray}
  (\Delta S_x)^2 & = &  4 - 4 a^2 (1  - \cos 2 \theta) , \\
  (\Delta S_y)^2 & = &  4 - 4 a^2 (1  + \cos 2 \theta) , \\
  (\Delta S_z)^2 & = & 8 a^2 .
\end{subeqnarray}
 This shows that Stokes unpolarized states can have
``hidden'' polarization properties that are not quantified by the
corresponding degree of polarization. As exemplified above, also
pure states can carry hidden polarization in quantum optics, which
is in contrast to classical optics.

When the Stokes parameters $\langle \hat{S}_x \rangle$, $\langle
\hat{S}_y \rangle$, and $\langle \hat{S}_z \rangle$ are all zero, it
also follows from relation (\ref{Eq:Svec2}) that Stokes unpolarized
$N$-photon states satisfy
\begin{equation}
  (\Delta S_x)^2 + (\Delta S_y)^2 + (\Delta S_z)^2 = N (N + 2) .
\end{equation}
The only states of the form (\ref{Eq:2phUnpol}) that have isotropic
fluctuations, i.e., satisfy $(\Delta S_x)^2 = (\Delta S_y)^2 =
(\Delta S_z)^2 = 8/3$, are seen to be those characterized by
$(a,\theta) = (1/\sqrt{3}, (2 m + 1) \pi/4)$, where $m \in \{ 0 , 1
, 2 , 3 \}$. These are equipartition states in the considered basis,
and consequently can be seen as relative-phase eigenstates
\cite{RelPh,PST}. We also note that the $x$-, $y$-, and
$z$-variances vanish for $(a,\theta) = (1/\sqrt{2},\pi/2)$,
$(a,\theta) = (1/\sqrt{2},0)$, and $a = 0$, respectively. Hence,
these states only have one or two nonvanishing components in the
horizontal-vertical Fock basis.

%%%%%%%%%%%%%%%%%%%%%%%%%%%%%%%%%%%%%%%%%%%%%%%%%
\begin{figure}
  \includegraphics[width=0.85\columnwidth]{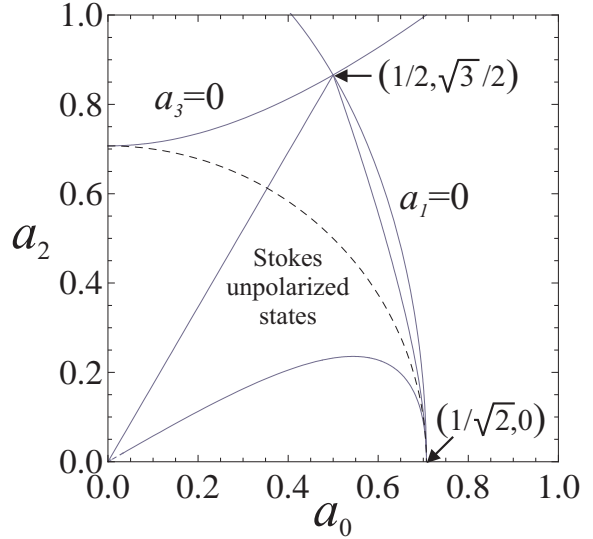}
  \caption{Permissable state probability amplitudes for pure three-photon states that are Stokes
 unpolarized.}
\label{Fig: unpolarized}
\end{figure}
%%%%%%%%%%%%%%%%%%%%%%%%%%%%%%%%%%%%%%%%%%%%%%%%

From the examples given so far, one may be led to believe that
Stokes unpolarized states have a symmetry with respect to
permutation of the horizontally and vertically polarized modes. This
is a chimera, however, as is seen in excitation manifold $N=3$. In
this manifold, there exists no unpolarized pure state with only one
nonzero probability amplitude or exactly one vanishing probability
amplitude. In order to find the unpolarized states, let us express
the probability amplitudes as $c_k = a_k \exp(i \theta_k)$, where
$a_k$ and $\theta_k$ are real, $a_k$ is non-negative, and we set
$\theta_0=0$ to remove an unimportant overall phase. From
Eqs.~(\ref{Eq:UnpolCondS12}) and (\ref{Eq:UnpolCondS3}), one can
then derive the relations
\begin{equation}
  a_1 = \frac{\sqrt{3 - 6 a_0^2 - 2 a_2^2}}{2} ,
  \qquad
  a_3 = \frac{\sqrt{1+2 a_0^2 - 2 a_2^2}}{2} .
  \label{Eq:a1a3}
\end{equation}
These equations can be used to delineate limits for $a_0$ and $a_2$
by looking at the values for which $a_1=0$ and $a_3=0$,
respectively. These limits are shown in Fig.~\ref{Fig: unpolarized},
where the axes represent the additional limits $a_0=0$ and $a_2=0$.
To satisfy Eq.~(\ref{Eq:UnpolCondS12}), one can view the terms
$\sqrt{3} a_0 a_1 \exp (-i \theta_1)$, $2 a_1 a_2 \exp [ i
(\theta_1-\theta_2) ] $, and $\sqrt{3} a_2 a_3 \exp [ i
(\theta_2-\theta_3)] $ as three vectors in the complex plane forming
a triangle when Eq.~(\ref{Eq:UnpolCondS12}) is fulfilled. This is
only possible if the triangle inequality is satisfied, i.e., the
length of any one vector cannot be larger than the sum of the
lengths of the remaining two vectors. The borders of these
inequalities can be written as
\begin{subeqnarray}
  a_2 & = & \sqrt{3} a_0 ,
  \slabel{eq:LeftBorder} \\
  a_2 & = & - \sqrt{3} a_0 - \sqrt{6} \cos  \left ( \frac{2 \pi + \arccos
    \sqrt{2} a_0}{3} \right ) ,
  \slabel{eq:RightBorder} \\
  a_2 & = & \sqrt{3} a_0 - \sqrt{6} \cos \left ( \frac{\pi + \arccos \sqrt{2}
    a_0}{3} \right )
\slabel{eq:LowerBorder} .
\end{subeqnarray}
In Fig.~\ref{Fig: unpolarized}, these borders form the innermost,
sail-shaped ``triangle'' with vertices $(0, 0)$, $(1/\sqrt{2}, 0)$,
and $(1/2, \sqrt{3}/2)$. This area comprises the allowed values for
$a_0$ and $a_2$, ultimately given by Eq.~(\ref{Eq:UnpolCondS12}).
For any permissible pair $(a_0,a_2)$, one can obtain the values of
$a_1$ and $a_3$ through the relations (\ref{Eq:a1a3}). Then one can
arbitrarily chose $\theta_1$ and subsequently find the four pairs of
values of $\theta_1-\theta_2$ and $\theta_2-\theta_3$ that make the
vectors corresponding to the three terms in
Eq.~(\ref{Eq:UnpolCondS12}) form a triangle. When one of the
triangle inequalities is exactly satisfied, i.e., when we are on one
of the borders of the sail-shaped area in Fig.~\ref{Fig:
  unpolarized}, there are only two solutions for $\theta_2$ and
$\theta_3$ once $\theta_1$ is chosen, namely $(\theta_2,\theta_3) = (2
\theta_1 + \pi,3 \theta_1 + \pi)$ and $(2 \theta_1 - \pi,3 \theta_1 -
\pi)$. For example, on the left border described by
Eq.~(\ref{eq:LeftBorder}), the states take the form
\begin{eqnarray}
  \lefteqn{a_0 \ket{0,3} + e^{i \theta_1}
  \sqrt{3 \left( \frac{1}{4} - a_0^2 \right)} \, \ket{1,2}} & & \nonumber \\
  & & \pm \left( e^{i 2 \theta_1} \sqrt{3} a_0 \ket{2,1} + e^{i 3
      \theta_1} \sqrt{\frac{1}{4} - a_0^2} \, \ket{3,0} \right) ,
  \label{eq:LeftBorderStates}
\end{eqnarray}
where $0 \leq a_0 \leq 1/2$.

For an arbitrary three-photon unpolarized pure state, the variances
of the Stokes operators are
\begin{subeqnarray}
  (\Delta S_{x,y})^2 & = & 3 +4 (a_1^2 + a_2^2) \nonumber \\
  & & \pm 4 \sqrt{3} \, \left[ a_0 a_2 \cos \theta_2 +
    a_1 a_3 \cos (\theta_3 - \theta_1) \right] \, , \nonumber \\ \\
  (\Delta S_z)^2 & = & 9 - 8 (a_1^2 + a_2^2) \, ,
  \slabel{eq:SzVar3ph}
\end{subeqnarray}
where $x$ and $y$ correspond to the plus and minus sign,
respectively. Equations (\ref{eq:SzVar3ph}) and (\ref{Eq:a1a3}) give
the curves
\begin{equation}
  a_2 = \sqrt{3 a_0^2 - \frac{(\Delta S_z)^2 - 3}{4}}
\end{equation}
on which $\Delta S_z$ is constant. In particular, the states (\ref{eq:LeftBorderStates})
correspond to $(\Delta S_z)^2 = 3$.

Using relations (\ref{Eq:a1a3}), it can also be verified that the
symmetry condition $a_k = a_{N-k}$ is equivalent to $a_0^2 + a_2^2 =
1/2$.  Hence, states whose probability amplitudes have this symmetry
are located on the dashed circle arc in
Fig.~\ref{Fig: unpolarized}, which has the vertex $(a_0,a_2) =
(1/\sqrt{2},0)$ as one of its end points.

We note that if $c_k = u_k$, $\forall k$, is a solution of
Eqs.~(\ref{Eq:UnpolCondS12})-(\ref{Eq:UnpolCondS3}), then $c_k =
u_{N-k}^\ast$, $\forall k$, is a solution too. In manifold $N = 3$,
any such pair of solutions that do not have the symmetry $a_k =
a_{N-k}$, will correspond to one point on each side of the dashed
circle arc. For example, the states corresponding to the two
vertices $(0,0)$ and $(1/2,\sqrt{3}/2)$ can be seen as such ``mirror
images'' of each other. It is clear from Fig.~\ref{Fig: unpolarized}
that the vertices of the sail-shaped area correspond to states with
only two nonvanishing components in the used basis. The states
corresponding to the mirror-image vertices are given by
Eq.~(\ref{eq:LeftBorderStates}), and the states corresponding to the
remaining vertex are $[\ket{0,3} + \exp (i \theta_3)
\ket{3,0}]/\sqrt{2}$, which indeed have the discussed symmetry.

Above, we have seen that states that are unpolarized according to the
Stokes definition can have anisotropic polarization fluctuations.
Perhaps a more dramatic example is demonstrated by the unpolarized
state $\ket{1,1}$ corresponding to $a = 0$ in
Eq.~(\ref{Eq:2phUnpol}). A rotation by 45 degrees around its axis of
propagation transforms this state into $[(\sqrt{2}+i)\ket{0,2} +
(\sqrt{2}-i)\ket{2,0}]/\sqrt{6}$, which is orthogonal to the original
$\ket{1,1}$~\cite{usachev}. Hence, despite being unpolarized, the
state $\ket{1,1}$ can be transformed into a perfectly distinguishable
state by a simple geometrical rotation. This is due to the fact that
this change cannot be detected by any linear combination of the Stokes
operators, as it requires higher-order field correlation
measurements. The classification of states according to Stokes degree
of polarization is hence insufficient already in excitation manifold
$N = 2$.

\section{Desiderata for a quantum degree of polarization}
\label{Sec:Desiderata}

\subsection{SU(2)-invariant quantum states}
\label{Sec:unpolarized}

As we have shown in the previous section, one is ill advised to
describe polarization properties of quantum fields by a direct analogy
with the classical description. A different starting point is needed.

In this respect, we recall that (linear) polarization
transformations are generated by the Stokes operators
(\ref{Stokop}). However, $\hat{S}_0$ induces only a common phase
shift to all the states in any given subspace, and below we will
argue that such phases do not change the polarization and can thus
be omitted.  Therefore, we restrict ourselves to the SU(2)
transformations, generated by $\hat{S}_{x}$, $\hat{S}_{y}$ and
$\hat{S}_{z}$. In fact, since each of these operators is
proportional to the commutator of the others, two generators
suffice. It is well known that $\hat{S}_{y}$ generates rotations
around the direction of propagation, whereas $\hat{S}_{z}$
represents differential phase shifts between the modes. Any
polarization transformation is thus photon-number preserving and can
be expressed as
\begin{equation}
  \label{Upol}
  \hat{U}_{\mathrm{pol}} (\alpha, \beta, \gamma) =
  e^{- i \alpha \hat{S}_{z}} e^{- i \beta \hat{S}_{y}} e^{- i \gamma \hat{S}_{z}} \, .
\end{equation}
This also means that they can be realized with linear optics.
Experimentally, birefringent plates in rotation mounts are the only
components needed, and consequently these transformations can be
simply and inexpensively achieved in a laboratory.

There is a consensus that the SU(2)-invariant states, which satisfy
$\hat{U}_{\mathrm{pol}} \hat{\sigma} \hat{U}_{\mathrm{pol}}^{\dagger} =
\hat{\sigma}$, are unpolarized. These states are known to
be of the form \cite{prakash,Girish,soderholm}:
\begin{equation}
  \label{eq:denunpol}
  \rhounpol =  \bigoplus_{N=0}^\infty \pi_N \, \hat{\sigma}_{N} \, ,
\end{equation}
where $\pi_N$ is the probability of finding the state $\rhounpol$ in
excitation manifold $N$, and $\hat{\sigma}_N$ is the only unpolarized
$N$-photon state
\begin{equation}
  \label{eq:N-phounp}
  \hat{\sigma}_{N} = \frac{1}{N+1} \hat{\openone}_{N} \, .
\end{equation}
Here, $\hat{\openone}_{N}$ is the projector onto the $N$-photon
subspace, namely
\begin{equation}
  \label{eq:1_N}
  \hat{\openone}_{N} = \sum_{k=0}^{N} \ket{k, N-k}
  \bra{k,N-k} \, .
\end{equation}
One notices that if a pure state is written as $\sum_{N=0}^\infty
c_{N,k}\ket{k, N - k}$, then coherence terms of the form
\begin{equation}
  \label{eq:1}
  c_{N,k} c^\ast_{N^{\prime},k^{\prime}}
  \ket{k, N - k} \bra{k^{\prime}, N^{\prime} - k^{\prime}}
\end{equation}
for $N \neq N^{\prime}$, can neither be induced nor measured by the
Stokes operators.  In consequence, $\hat{\sigma}$ appears as a direct
sum over the excitation manifolds in Eq.~(\ref{eq:denunpol}) and any
common phase to all the states in any given excitation manifold is
inconsequential for any polarization characteristics.

\subsection{Requirements for polarization measures}
\label{Sec:criteria}

Before discussing specific quantum measures of polarization, it is
worthwhile to look at requirements and desiderata for such measures.

\textit{Requirement 1}.  A first requirement for any reasonable degree
of polarization $\pgen$ is
\begin{equation}
  \pgen(\density) = 0 \quad
  \Leftrightarrow \quad \density \ \textrm{is unpolarized} \, .
  \label{Eq: req1}
\end{equation}
This immediately
rules out the possibility of defining the degree of polarization as
a function of the purity $\Tr (\hat{\varrho}^{2})$. The state
$\ket{0,0}$ is pure and unpolarized, while a two-mode thermal state
(with the same mean photon number in each mode) is maximally mixed
(under the constraint of a fixed average number of photon number)
and likewise unpolarized. Also, any state $p_0 \proj{0,0}{0,0} + p_N
\hat{\openone}_{N}/(N+1)$, where $p_0$ and $p_N$ are both
nonvanishing, is unpolarized and mixed, but not maximally mixed
(under the same constraint). Hence, unpolarized quantum states span
the whole purity scale.

\textit{Requirement 2}. A second requirement is SU(2) invariance
\cite{luis review}
\begin{equation}
  \pgen(\density) =
  \pgen   (\hat{U}_{\mathrm{pol}} \, \hat{\varrho} \,
  \hat{U}_{\mathrm{pol}}^\dagger) \, .
  \label{Eq:req3}
\end{equation}
Hence, the measure is invariant under polarization transformations.
For instance, the Stokes degree of polarization (6) fulfills this
condition.

\textit{Requirement 3}. A third requirement that has been put
forward is that the measure should not depend on the coherences
between different manifolds~\cite{luis review}. The basis for this
requirement is that since $\hat{S}_{0}$ commutes with all Stokes
operators, a polarization measurement (a measurement of any linear
combination of the Stokes operators) on an arbitrary state
\begin{equation}
  \density = \sum_{N, N^\prime=0}^\infty \sum_{k=0}^{N}
  \sum_{k^\prime = 0}^{N^\prime} \varrho_{Nk, N^{\prime} k^{\prime}}
\ket{k, N- k} \bra{k^\prime, N^\prime - k^\prime} \, , \label{Eq:
general density}
\end{equation}
does not on average alter the photon-number distribution
\begin{equation}
  p_N = \sum_{k=0}^{N} \varrho_{Nk,Nk} \,   \label{6}
\end{equation}
and the measurement outcome will not depend on any coherences between
the manifolds.

On the other hand, a von Neumann measurement of the number of
photons gives an outcome $N$ with probability $p_N$ and, at the same
time, the state $\hat{\varrho}$ collapses into the $N$-photon state
\begin{equation}
  \hat{\varrho}_N =
  \frac{1}{p_N}\sum_{k, k^{\prime}=0}^N
  \varrho_{Nk,Nk^{\prime}} |k , N- k \rangle
  \langle k^{\prime}, N - k^{\prime}| \, .
  \label{7}
\end{equation}
Considering all possible outcomes, we obtain the block-diagonal
state
\begin{equation}
  \mathcal{B} [\hat{\varrho}] = \bigoplus_{N=0}^{\infty} p_N \, \hat{\varrho}_N \,
  , \label{8}
\end{equation}
where the ideal non-selective measurement of the total photon number
is described by the map
\begin{equation}
  \mathcal{B} : \hat{\varrho} \mapsto
\sum_{N=0}^{\infty}  \hat{\openone}_{N} \, \hat{\varrho} \,
\hat{\openone}_{N} \, . \label{10}
\end{equation}
This is a quantum channel~\cite{Nie00} preserving both the
polarization properties and the photon-number distribution of the
state $\hat{\varrho}$, and provides an operational meaning for the
channel. Alternatively, the map $\mathcal{B}$ can be viewed as
randomization of the phases between superpositions of
states in different excitation manifolds. Using this map, requirement 3
can be expressed as
\begin{equation}
  \pgen   ( \hat{\varrho} ) =
  \pgen   ( \mathcal{B} [\hat{\varrho}]) \, . \label{Eq:req3b}
\end{equation}

Polarization measures that depend on coherences between different
manifolds but fulfill requirement 2 can be made to fulfill
Eq.~(\ref{Eq:req3b}) by applying the measure to the channel output as
will be done below in Eq.~(\ref{Eq: block diag}).

Notice that some polarization-measure candidates (such as the entropy
$\mathbb{S}$) are only positive semidefinite, so that $0 \leq
\mathbb{S} (\density) < \infty$. In this case, a common ``remedy'' is
to normalize the measure through the transformation $\pgen =
\mathbb{S} / (1 + \mathbb{S})$, which guarantees the
condition
\begin{equation}
  0 \leq \pgen(\density) \leq 1 \, .
  \label{Eq: req2}
\end{equation}
Such a rescaling keeps the ``ordering'' of states intact.  Indeed, the
induced ordering of states is more important than the numerical value,
especially when the measure does not have a clear operational meaning.

The requirements 1-3 can be supplemented by a number of desiderata.  The
most common, in particular among experimentalists, is that $\pgen$
should be operational and easily measurable.  Theoreticians, on the
other hand, desire that the measure is easy to compute. Unfortunately,
in general, these wishes are conflicting.

From an experimental point of view, the measure may favor a number of
different operational characteristics.  One could, e.g., quantify the
maximum visibility achievable in a polarization interference
measurement~\cite{bjork}.  Such a measure would fulfill all three
requirements and would also have a direct operational meaning, but it
would not be easily measurable, in general, as one would not know what
are the polarization transformations that yield the maximum and
minimum interference intensity.

One could alternatively determine how close a given state is to a
polarization minimum uncertainty state~\cite{luisQ}. Such a measure
would also fulfill the requirements and have a relatively clear
operational meaning, but it would require polarization tomography, a
complicated measurement procedure, to be determined.

Another possibility is to evaluate the polarization fluctuations. In
this case, it would probably make sense to assess the fluctuations
along the polarization coordinate that gives the smallest
fluctuations. This would give an idea about the smallest detectable
polarization transformations and hence have an operational
meaning. However, for a general state, the measure would be difficult
to determine and, in general, also difficult to compute.

In summary, we see that there are many possibilities of defining a
measure of polarization. We have argued that our three requirements
are reasonable conditions for any such a measure. Note that all are
closely related to the properties of the Stokes operators, which we
have taken as our starting point. Any ensuing degree of polarization
will have their particular merits and drawbacks.

\section{Quantum degrees of polarization}
\label{Sec:Examples}

\subsection{Distance-based measures}
\label{Sec:distance based}

After our discussion in Section~\ref{Sec:unpolarized}, it seems sensible
to define the degree of polarization as the shortest distance between
the considered state and the set $\mathcal{U}$ of unpolarized states
$\hat{\sigma}$ given in Eq.~(\ref{eq:denunpol}). Similar notions have
been successfully applied to other key concepts such as
nonclassicality~\cite{Hillery,Victor,Marian},
entanglement~\cite{vedral} and quantum
information~\cite{Schumacher,Caves,Preskill,Nielsen}.

Several distance measures have been proposed, such as the
Hilbert-Schmidt and Bures distances~\cite{klimov}. We also include
the Chernoff distance, recently used to quantify the nonclassicality
of Gaussian states~\cite{boca} and polarization~\cite{ghiu}. For an
arbitrary state $\density$, these measures are given by
\begin{subeqnarray}
  \slabel{Eq: defPHS}
  \phs (\density) & = & \inf_{\rhounpol \in \mathcal{U}} \
  \tr [( \density - \rhounpol)^2 ] , \\
  \slabel{Eq: defPB}
  \pb (\density) & = & 1 - \sup_{\rhounpol \in \mathcal{U}}
  \sqrt{F (\density,\rhounpol)} \, , \\
  \slabel{Eq: defPChern}
  \pc (\density) & = & 1 - \sup_{\rhounpol \in \mathcal{U}} \left [
    \inf_{s \in [0,1]} \tr (\density^s \rhounpol^{1-s}) \right ]\, ,
\end{subeqnarray}
where the infimum in  Eq.~(\ref{Eq: defPChern}) is taken over a
function that is continuous with respect to $s$ \cite{hiai}, and the fidelity is
\begin{equation}
  F (\density,\rhounpol) = \{\tr [(\rhounpol^{1/2}
  \density \rhounpol^{1/2})^{1/2}]\}^2 \, .
\end{equation}

While all these definitions seem sensible, they do not satisfy
requirement 3; that is, they are sensitive to coherences between
different excitation manifolds \cite{luis review}. To bypass this drawback, we apply
requirement 3, i.e., we replace the states by the corresponding
block-diagonal density matrices:
\begin{equation}
\label{Eq: block diag} \mathbb{P}_{Z \mathrm{b}} (\hat{\varrho}) =
\mathbb{P}_Z (\mathcal{B} [\hat{\varrho}]) , \quad Z \in \{
\mathrm{HS} , \mathrm{B} , \mathrm{C} \} .
\end{equation}
These measures can thus be seen as applying the original measures on
the block-diagonal output state of the photon-number measurement
channel (\ref{10}), whose input state is $\hat{\varrho}$. Using the
fact that $\mathcal{B} [\hat{\varrho}]$ and $\hat{\sigma}$ commute,
we find the following general formulas:
\begin{subeqnarray}
  \phsb (\hat{\varrho}) & = & \sum_{N=0}^\infty p_N^2 \left (
    \xi_N^{(2)} -\frac{1}{N+1} \right )
  \slabel{Eq: PHSDiag}, \\
  \pbb (\hat{\varrho}) & = & 1 - \left [ \sum_{N=0}^\infty
    \frac{p_N}{N+1} \left (\xi_N^{(1/2)} \right )^{2} \right ]^{1/2} ,
  \slabel{Eq: PBDiag} \\
  \pcb (\hat{\varrho}) & = & 1 - \inf_{s \in [0,1]} \left[
    \sum_{N=0}^\infty p_N (N+1)^{1-1/s} \left( \xi_N^{(s)}\right)
    ^{1/s}\right]^{s} , \nonumber \\ \slabel{Eq: PchernDiag}
\end{subeqnarray}
where $\eigen_{N,n}$ are the eigenvalues of $\hat{\varrho}_N$ and
$\xi_N^{(s)} \equiv \sum_{n=0}^N\eigen_{N,n}^s$. These measures
fulfill our three requirements for a degree of polarization.
Obviously, $\pbb (\hat{\varrho}) \leq \pcb (\hat{\varrho})$.

Any pure state $\hat{\varrho} = | \Psi \rangle \langle \Psi |$
satisfies $\tr (\density_N^2) = 1$ in each manifold with nonzero
excitation probability. Hence, for any such $N$, one of the
eigenvalues $\eigen_{N,n}$ equals unity and the rest of them vanish.
The degrees of polarization are thus given by
\begin{subeqnarray}
  \phsb (| \Psi \rangle) & = & \sum_{N=0}^\infty
  p_N^2 \frac{N}{N+1} , \\
  \pbb (| \Psi \rangle) & = & 1 -
  \left [ \sum_{N=0}^\infty \frac{p_N}{N+1}\right ]^{1/2} ,  \\
  \pcb (| \Psi \rangle) & = & 1 - \inf_{s \in [0,1]}\left[
    \sum_{N=0}^\infty p_N(N+1)^{1-1/s}\right] ^s .
\end{subeqnarray}
These expressions involve only the excitation probabilities $p_N$.
Thus, for pure states, the block-diagonal distance degrees of
polarization are insensitive to the form(s) of $\hat{\varrho}_N$. In
the special case of a pure $N$-photon state $\hat{\varrho} =
|\Psi_{N}\rangle \langle \Psi_{N}|$, the above expressions simplify
to $\mathbb{P}_{\mathrm{HSb}} (|\Psi_{N}\rangle) =
\mathbb{P}_{\mathrm{Cb}} (|\Psi_{N}\rangle) = N/(N+1)$ and
$\mathbb{P}_{\mathrm{Bb}}(|\Psi_{N}\rangle) = 1 - (N+1)^{-1/2}$. All
of them tend to unity for large $N$.

It is clear that for a fixed excitation manifold $N$, pure states
have a higher degree of polarization than any mixed state, as one
would intuitively expect. Using Lagrange multipliers and numerical
optimization, we have derived the block-diagonal
states that for a given average photon number $\bar{N}$ have the highest degrees of
polarization \cite{unpublished}. For the Hilbert-Schmidt measure,
these maximally polarized states are of the form
\begin{equation}
  \frac{\lceil \bar{N} \rceil - \bar{N}}{\lceil \bar{N} \rceil} \, \proj{0,0}{0,0}
  + \frac{\bar{N}}{\lceil \bar{N} \rceil} \, \proj{\Psi_{\lceil \bar{N} \rceil}}{\Psi_{\lceil \bar{N} \rceil}}
\end{equation}
if $\bar{N} \geq \sqrt{\lfloor \bar{N} \rfloor (\lfloor \bar{N}
\rfloor + 2)}$, and of the form
\begin{equation}
  \lim_{M \rightarrow \infty} \left( \frac{M - \bar{N}}{M - \lfloor \bar{N} \rfloor} \,
  \proj{\Psi_{\lfloor \bar{N} \rfloor}}{\Psi_{\lfloor \bar{N} \rfloor}} +
  \frac{\bar{N} - \lfloor \bar{N} \rfloor}{M - \lfloor \bar{N} \rfloor} \, \proj{\Psi_{M}}{\Psi_{M}} \right) \, ,
\end{equation}
if $\bar{N} \leq \sqrt{\lfloor \bar{N} \rfloor (\lfloor \bar{N}
\rfloor + 2)}$. Here, $\lceil \bar{N} \rceil$ denotes the smallest
integer larger than or equal to $\bar{N}$, whereas $\lfloor \bar{N}
\rfloor$ is the largest integer smaller than or equal to $\bar{N}$.
Hence, the maximal polarization degree is given by the somewhat
``rounded'' staircase function
\begin{equation}
\mathbb{P}_\mathrm{HSb}^\mathrm{max} = \left\{ \frac{\lfloor \bar{N}
\rfloor}{\lfloor \bar{N} \rfloor + 1} , \quad \bar{N} \leq
\sqrt{\lfloor \bar{N} \rfloor (\lfloor \bar{N} \rfloor + 2)} , \atop
\frac{\bar{N}^2}{\lceil \bar{N} \rceil (\lceil \bar{N} \rceil + 1)}
, \quad \bar{N} \geq \sqrt{\lfloor \bar{N} \rfloor (\lfloor \bar{N}
\rfloor + 2)} . \right.
\end{equation}

%%%%%%%%%%%%%%%%%%%%%%%%%%%%%%%%%%%%%%%%%%%%%%%%%
\begin{figure}
  \includegraphics[width=0.85\columnwidth]{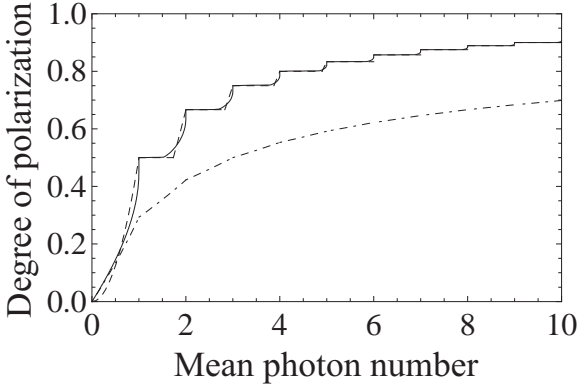}
  \caption{The maximal degree of polarization vs. the average photon
    number for the block-diagonal Hilbert-Schmidt (dashed), Bures (dash-dotted),
    and Chernoff measures (solid).}
  \label{Fig: maximal}
\end{figure}
%%%%%%%%%%%%%%%%%%%%%%%%%%%%%%%%%%%%%%%%%%%%%%%%

If instead the Bures or Chernoff measure is used, the maximally
polarized states are given by
\begin{equation}
  \left( \lceil \bar{N} \rceil - \bar{N} \right)
  \proj{\Psi_{\lceil \bar{N} \rceil - 1}}{\Psi_{\lceil \bar{N} \rceil - 1}} +
  \left( \bar{N} + 1 - \lceil \bar{N} \rceil \right)
  \proj{\Psi_{\lceil \bar{N} \rceil}}{\Psi_{\lceil \bar{N} \rceil}}
\end{equation}
and the corresponding degrees of polarization are
\begin{eqnarray}
  \mathbb{P}_\mathrm{Bb}^\mathrm{max} & = & 1 -
  \sqrt{\frac{2 \lceil \bar{N} \rceil - \bar{N}}{\lceil \bar{N} \rceil (\lceil \bar{N} \rceil +
  1)}} , \\
  \mathbb{P}_\mathrm{Cb}^\mathrm{max} & = & 1 -
  \inf_{s \in [0,1]} \left[
  \lceil \bar{N} \rceil^\frac{s-1}{s} (\lceil \bar{N} \rceil - \bar{N}) \right. \nonumber \\
  & & \left. + (\lceil \bar{N} \rceil + 1)^\frac{s-1}{s} (1 + \bar{N}
  - \lceil \bar{N} \rceil) \right]^{s} .
\end{eqnarray}
Plots of $\mathbb{P}_\mathrm{HSb}^\mathrm{max}$,
$\mathbb{P}_\mathrm{Bb}^\mathrm{max}$, and
$\mathbb{P}_\mathrm{Cb}^\mathrm{max}$ are shown in Fig.~\ref{Fig:
maximal}. We note that any pure $N$-photon state $\ket{\Psi_{N}}$ is
maximally polarized according to any of these three measures. We
also note that $\mathbb{P}_\mathrm{Bb}^\mathrm{max}$ has a strictly
positive derivative with respect to $\bar{N}$, whereas
$\mathbb{P}_\mathrm{HSb}^\mathrm{max}$ and
$\mathbb{P}_\mathrm{Cb}^\mathrm{max}$ have non-negative but
discontinuous derivatives.

\subsection{Other quantum polarization measures}
\label{Sec:other measures}

Several other quantum degrees of polarization have been proposed. One
of them is based on the SU(2) $Q$ function~\cite{luisQ}, which is
defined as
\begin{equation}
  Q_{\density} ( \bm{\Omega} )  =  \sum_{N=0}^\infty
  \frac{N+1}{4 \pi} \bra{N;\bm{\Omega}} \density \ket{N;\bm{\Omega}} =
  Q_{\mathcal{B} [\density]} ( \bm{\Omega} )  \, .
\end{equation}
Here, $\bm{\Omega}= (\vartheta, \varphi)$ and $\vartheta$ and
$\varphi$ are the polar and azimuthal angles over the unit 2-sphere
$S^2$, and $\ket{N;\bm{\Omega}}$ are the $N$-photon SU(2) coherent
states
\begin{equation}
  \ket{N;\bm{\Omega}} = \sum_{k=0}^N
  \left (
    \begin{array}{c}
      N \\ k
    \end{array}
  \right )^{1/2}
  \left ( \cos \frac{\vartheta}{2} \right )^k
  \left ( \sin \frac{\vartheta}{2} \right )^{N-k} e^{i k \varphi}
  \ket{k,N-k} .
\end{equation}
For any unpolarized state (\ref{eq:denunpol}), the $Q$ function takes
the constant value $(4 \pi)^{-1}$.  Apart from the unpolarized vacuum
state, any SU(2) coherent state has a $Q$ function that is highly
peaked around some angle $\bm{\Omega}_0$. For example, for a SU(2)
coherent state centered around $\vartheta=0$, that is, the state
$\ket{N; 0}$, we have
\begin{equation}
  Q_{\ket{N; 0}} ( \bm{\Omega} )= \frac{N+1}{4 \pi}
  \left ( \cos \frac{\vartheta}{2} \right )^{2 N} .
\end{equation}
The idea behind a $Q$ function-based measure is to assess the spread of $Q$ over the
sphere by comparing with a uniform distribution:
\begin{eqnarray}
  D_{Q} (\density) & =  & 4 \pi \int \left [
    Q_{\density} (\bm{\Omega} ) - \frac{1}{4 \pi}\right ]^2
  \mathrm{d} \bm{\Omega} \nonumber \\
 & = &  4 \pi \int
   Q_{\hat{\varrho}}^2(\bm{\Omega}) \, \mathrm{d} \bm{\Omega} - 1 .
\end{eqnarray}
However, since $ 0 \leq D_{Q} (\density) < \infty$, the associated
degree of polarization is defined as
\begin{equation}
  \pq (\density) = \pq (\mathcal{B} [\density])= \frac{D_{Q} (\density)}
  {D_{Q} (\density) + 1} .
\end{equation}
This degree favors polarization minimum uncertainty states as it
measures the ``area'' of $Q$, but is insensitive to its shape or
orientation. The measure can be obtained experimentally, but only
through rather involved polarization tomography.

%%%%%%%%%%%%%%%%%%%%%%%%%%%%%%%%%%%%%%%%%%%%%%%%%
\begin{figure}
 \includegraphics[width=0.85\columnwidth]{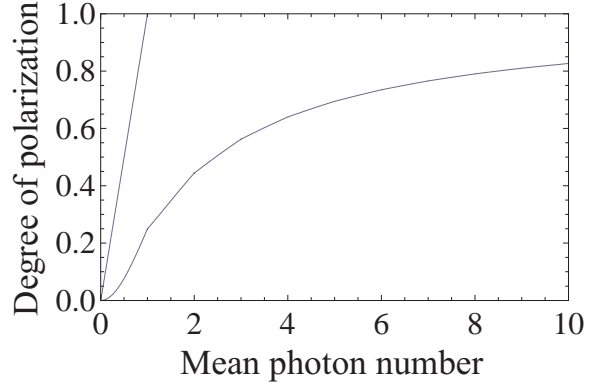}
  \caption{The maximal degree of polarization vs. the average photon
    number for the $Q$ function-based measure $\pq$ (solid line),
    the measure based on SU(2)-induced distinguishability $\pd$, and the
    purity-based measure $\pp$ (dashed coincident lines).}
  \label{Fig: maximal Q}
\end{figure}
%%%%%%%%%%%%%%%%%%%%%%%%%%%%%%%%%%%%%%%%%%%%%%%%

For an SU(2) coherent state, $\pq (\ket{N;\bm{\Omega}}) =
[N/(N+1)]^2$.  As the SU(2) coherent states are polarization minimum
uncertainty states they are maximally polarized according to the
definition of $\pq$. For an average photon number $\bar{N}$, the
superposition, or mixture, of the SU(2) coherent states
$\ket{n-1;\bm{\Omega}}$ and $\ket{n;\bm{\Omega}}$, with
probabilities $p_{n-1} = n - \bar{N}$ and $p_{n} = 1 + \bar{N} - n$,
respectively, are the states with the maximal $\pq$-degree of
polarization.

Another proposed polarization measure is given in~\cite{bjork}.
Using Tr$(\hat{\varrho}_1 \hat{\varrho}_2)$ as the overlap for mixed
states, the definition for pure states in Ref.~\cite{bjork} is
generalized to
\begin{eqnarray}
  \pd (\density) & = & \pd (\mathcal{B} [\density]) \nonumber \\
  & = & \left [1 - \inf_{\upol}\sum_{N=0}^\infty p_N
    \tr(\density_N \upol \density_N \upol^{\dagger})
  \right ]^{1/2} .
\end{eqnarray}
This definition is based on the probability averaged minimal overlap
between a state and all of its SU(2) transformed states. Hence, it
gives the (square root of the) maximum visibility one can achieve by
using a polarization interferometer.  The problem in this case is to
find and implement the polarization projection and the subsequent
polarization transformation that achieves the maximum polarimetric
visibility. In contrast to all the previous measures in this
section, $\mathbb{P}_\mathrm{d}$ may assign the degree of
polarization unity for states with a finite average excitation. It
has been shown that any pure state having an odd photon number is
maximally polarized in the sense that $\pd=1$~\cite{sehat}. One may
conjecture that pure states with an even number of photons
(excluding the vacuum state) also are maximally polarized according
to this definition, but to the best of our knowledge no proof
thereof exists, except for $N = 2$.

The last speculation makes it tempting to define a degree of
polarization in terms of the state purities in every excitation
manifold, as follows:
\begin{equation}
  \pp (\density)= \pp (\mathcal{B} [\density]) =
  \sum_{N=1}^\infty p_N \frac{(N+1) \tr(\density_N^2)-1}{N},
\end{equation}
where we need the additional definition $\pp(\ket{0,0}) \equiv 0$.
Again, the maximally polarized states are the pure states in any
excitation manifold or any mixture, or superposition, thereof. The
measure makes no direct use of the Stokes operators (except for
being a direct sum over manifolds). This indicates that the measure
quantifies a distinguishability under a general energy-preserving
unitary transformation rather than the distinguishability under the
more restrictive unitary polarization transformations $\upol$.
Should the conjecture in the previous paragraph prove false, this
measure seems questionable.  A measurement will unfortunately be
difficult since the purity essentially must be assessed through
polarization tomography. In Fig.~\ref{Fig: maximal Q}, the maximum
degree of polarization for the measures $\pq$, $\pd$, and $\pp$ are
plotted.

\section{Nonlinear polarization transformations}
\label{Sec: nonlinear}

Above we have defined the set of ``proper'' polarization
transformations in Eq.~(\ref{Upol}) as all linear transformations
generated by the Stokes operators. Such a viewpoint has a basis both
in classical and quantum optics. Of course, one could think in more
general terms and allow nonlinear (energy-pre\-ser\-ving)
transformations, which can be represented as
\begin{equation}
  \label{eq:UNL}
  \hat{U}_{\mathrm{nl}} = e^{-i g(\hat{S}_{0}, \hat{\mathbf{S}})} \, ,
\end{equation}
where $g$ is an arbitrary nonlinear function. Such a set of
transformations includes a variety of effects such as polarization
squeezing \cite{squeezing} and excitation manifold-dependent
transformations. However, one could argue that both polarization
squeezing and manifold-dependent transformations can change a
state's degree of polarization, as we shall give an example of
below. Another reason for excluding such transformations is that
they are very difficult to implement experimentally.

If one allows nonlinear polarization transformations, $\pq$ will no
longer fulfill Eq.~(\ref{Eq:req3}). For example, the state
$\ket{\Psi_1} = ( \ket{N,0} + \ket{N^{\prime},0})/\sqrt{2}$, where
$N,N^{\prime} \neq 0$ has a $Q$ function that is concentrated on the north
pole and whose dispersion is
\begin{eqnarray}
  D_{Q} (\ket{\Psi_1})  & = & \frac{1}{4}
  \left [  \frac{ (N + 1)^2}{2N+1} +
  \frac{( N^{\prime}  + 1)^2}{2 N^{\prime} + 1} \right.
  \nonumber \\
  & &  \left. +
    2 \frac{(N + 1)(N^{\prime} + 1)}{N + N^{\prime} + 1}   \right ]  - 1 \, .
\end{eqnarray}
With a nonlinear polarization transformation it is possible to
transform the state $\ket{N^{\prime},0}$ to $\ket{0,N^{\prime}}$,
that is, to rotate this state to the south pole of the
representation sphere without affecting the state $\ket{N,0}$.
However, the state
$\ket{\Psi_2}=(\ket{N,0}+\ket{0,N^{\prime}})/\sqrt{2}$ has the
dispersion
\begin{eqnarray}
  \lefteqn{D_{Q} (\ket{\Psi_2})  = } & & \nonumber \\
  & &  \frac{1}{4} \left [
    \frac{(N + 1)^2}{2 N + 1} + \frac{(N^{\prime}+1)^2}{2N^{\prime}+1}
    +   2 \frac{(N + 1)! (N^{\prime} + 1)!}{(N + N^{\prime} + 1)!} \right ] - 1 \, , \nonumber \\
\end{eqnarray}
which is close to half the value of $D_{Q} (\ket{\Psi_1})$ when $N
\approx N^{\prime} \gg 1$. Hence, $\pq (\ket{\Psi_1}) > \pq
(\ket{\Psi_2})$.

Perhaps, superior future technology will make it natural to view also
nonlinear Stokes operator induced transformations as ``proper''
polarization transformations, in contrast to our definition. Such a
view would distance the quantum description of polarization even
further from the classical one.

\section{Discussion and conclusions}
\label{Sec:discussion conclusions}

As we have seen, defining a quantitative measure of polarization for
quantum fields is a task without any obvious or unique solution.
As a consequence, no universally accepted view on how to quantify
the polarization of such fields exists, and the prospects of this
happening seem bleak.

In this paper, we have advocated the view that the Stokes operators
should be central to any quantum polarization theory. Three of the
central requirements for a quantitative degree of polarization are
based on their properties. Adhering to this view would ascertain at
least partial correspondence between classical and quantum concepts
and descriptions of polarization. A consequence of this view is the
definition of a polarization transformation. This is a unitary
transformation generated by any linear combination of the Stokes
operators. Another rather unavoidable consequence is the definition of
an unpolarized quantum state as a state where each excitation manifold
is invariant under any polarization transformation.

It is clear from our discussion of maximally polarized states that
the proposed measures order the degree of polarization for states
differently. This is to be expected since each measure focuses on
one specific polarization property. For example, a state that can
become self-orthogonal under a polarization transformation, thus
having $\pd=1$, may not be even close to a polarization minimum
uncertainty state, which are states for which $\pq$ is large. We
therefore conjecture that different degrees of polarization will
coexist, and that they will find applications in different
polarization contexts.

We have also shown in Section~\ref{Sec: nonlinear} that allowing
nonlinear transformations as ``proper'' polarization transformations
will lead to profound differences in the way we view quantum
polarization. As long as such transformations are essentially
outside the realm of what is experimentally realizable for few
photon states, it seems reasonable to stick with the set of linear
transformations.

We have only discussed polarization properties for two-mode fields. In
principle, the formalism above will apply to any two harmonic
oscillators, but if we want to retain some connection to the classical
concepts of polarization, the two modes should be monochromatic,
co-propagating, approximately plane waves in approximately the same
temporal modes. Of course one could, in analogy with the development
in classical optics, start to define polarization concepts and degrees
of polarization for three-dimensional fields \cite{carozzi,setala,luis 3D}
(e.g., in strongly focused beams of light), or for
polarization-entangled, four-mode states. Attempts in this direction
have been made~\cite{karassiov,rubin,bushev,karassiov2}.  However, such generalizations of the
basic concepts are often difficult to interpret and to give an
operational meaning. Hence, it is probably more fruitful to refer to
such general multimode characteristics as field- or mode-correlations,
without using the word polarization.

\section*{Acknowledgements}

This work was supported by the Swedish Research Council (VR) and the
Swedish Foundation for International Cooperation in Research and
Higher Education (STINT). In addition, LLSS was supported by the
Spanish Research Directorate (Grant No FIS2008-04356), ABK was
supported by Consejo Nacional de Ciencia y Tecnolog\'{\i}a (Grant No
45704), and IG, PM, and TAM were supported by the Romanian Ministry
of Education and Research (Grant No IDEI-995/2007).


\begin{thebibliography}{00}

\bibitem{Sto52}
G.~G.~Stokes,
Trans. Cambridge Philos. Soc. 9 (1852) 399.

\bibitem{schott}
J.~R.~Schott,
Fundamentals of Polarimetric Remote Sensing,
SPIE, Bellingham,  2009.

\bibitem{barron}
L.~D.~Barron,
 Molecular Light Scattering and Optical Activity,
Cambridge University Press, Cambridge, UK, 2004.

\bibitem{azzam}
R.~M.~A.~Azzam, N.~M.~Bashara,
Ellipsometry and Polarized Light,
Elsevier, Amsterdam, 1987.

\bibitem{werner}
S.~Werner, O.~Rudow, C.~Mihalcea, E.~Oesterschulze,
Appl. Phys. A 66 (1998) S367.

\bibitem{klyshko}
D.~N.~Klyshko,
Phys. Lett. A 163 (1992) 349.

\bibitem{klyshko1997}
D.~N.~Klyshko,
Sov. Phys. JETP 84 (1997) 1065.

\bibitem{alojants}
A.~P.~Alodjants, S.~M.~Arakelian, A.~S.~Chirkin,
Quant. Semiclass. Opt. 9 (1997) 311.

\bibitem{luis2000}
A.~Luis, L.~L.~S\'anchez-Soto,
in: E.~Wolf (Ed.), Progress in Optics, vol. 41,
Elsevier, Amsterdam, 2000, p. 421.

\bibitem{klimov}
A.~B.~Klimov, L.~L.~S\'anchez-Soto, E.~C.~Yustas,
J.~S\"{o}derholm, G.~Bj\"{o}rk,
Phys. Rev. A 72 (2005) 033813.

\bibitem{luis review}
A.~Luis,
Opt. Commun. 273 (2007) 173.

\bibitem{Charly}
C.~H.~Bennett, F.~Bessette, G.~Brassard, L.~Salvail, J.~Smolin,
J. Cryptology 5 (1992) 3.

\bibitem{muller}
A.~Muller, J.~Breguet, N.~Gisin,
Europhys. Lett. 23 (1993) 383.

\bibitem{mattle}
K.~Mattle, H.~Weinfurter, P.~G.~Kwiat, A.~Zeilinger,
Phys. Rev. Lett. 76 (1996) 4656.

\bibitem{kwiat}
P.~G.~Kwiat, K.~Mattle, H.~Weinfurter,
A.~Zeilinger, A.~V.~Sergienko, Y.~Shih,
Phys. Rev. Lett. 75 (1995) 4337.

\bibitem{bouwmeester}
D.~Bouwmeester, J.-W.~Pan, K.~Mattle, M.~Eibl, H.~Weinfurter, A.~Zeilinger, Nature 390 (1997) 575.

\bibitem{francesco}
M.~Barbieri, F.~De Martini, G.~Di Nepi, P.~Mataloni,
G.~M.~D'Ariano, C.~Macchiavello,
Phys. Rev. Lett. 91 (2003) 227901.

\bibitem{radmark}
M.~R{\aa}dmark, M.~Zukowski, M.~Bourennane, New J. Phys. 11 (2009)
103016.

\bibitem{resch}
K.~J.~Resch, K.~L.~Pregnell, R.~Prevedel, A.~Gilchrist,
G.~J.~Pryde, J.~L.~O'Brien, A.~G.~White,
Phys. Rev. Lett. 98 (2007) 223601.

\bibitem{collett}
E.~Collett,
Am. J. Phys. 38 (1970) 563.

\bibitem{Bro1998}
C.~Brosseau,
Fundamentals of Polarized Light: A Statistical Optics Approach,
Wiley, New York, 1998.

\bibitem{RelPh}
A.~Luis, L.~L.~S\'anchez-Soto,
Phys. Rev. A 48 (1993) 4702.

\bibitem{PST}
G.~Bj\"{o}rk, J.~S\"{o}derholm, A.~Trifonov, T.~Tsegaye,
Phys. Scr. T102 (2002) 133.

\bibitem{usachev}
P.~Usachev, J.~S\"{o}derholm, G.~Bj\"{o}rk, A.~Trifonov,
Opt. Commun. 193 (2001) 161.

\bibitem{prakash}
H.~Prakash, N.~Chandra, Phys. Rev. A 4 (1971) 796

\bibitem{Girish}
G.~S.~Agarwal, Lett. Nuovo Cimento 1 (1971) 53.

\bibitem{soderholm}
J.~S\"{o}derholm, G.~Bj\"{o}rk, A.~Trifonov,
Opt. Spectrosc. (USSR) 91 (2001) 532.

\bibitem{Nie00}
M.~ A.~Nielsen, I.~L.~Chuang, Quantum Computation and Quantum
Information, Cambridge University Press, Cambridge, UK, 2000.

\bibitem{bjork}
G.~Bj\"{o}rk, J.~S\"{o}derholm, A.~Trifonov,
P.~Usachev, L.~L.~S\'anchez-Soto, A.~B.~Klimov,
Proc. SPIE 4750 (2002) 1.

\bibitem{luisQ}
A.~Luis, Phys. Rev. A 66 (2002) 013806; 71 (2005) 023810.

\bibitem{Hillery}
M.~Hillery,
Phys. Rev. A  35 (1987) 725.

\bibitem{Victor}
V.~V.~Dodonov, O.~V.~Manko, V.~I.~Manko, A.~W\"{u}nsche,
J. Mod. Opt. 47 (2000) 633.

\bibitem{Marian}
P.~Marian, T.~A. Marian, H.~Scutaru,
Phys. Rev. Lett. 88 (2002) 153601.

\bibitem{vedral}
V.~Vedral, M.~B.~Plenio, M.~A.~Ripping, P.~L.~Knight,
Phys. Rev. Lett. 78 (1997) 2275.

\bibitem{Schumacher}
B.~Schumacher, Phys. Rev. A 51 (1995) 2738.

\bibitem{Caves}
R.~Schack, C.~M.~Caves, Phys. Rev. A 60 (1999) 4354.

\bibitem{Preskill}
A.~M.~Childs, J.~Preskill, J.~Renes, J. Mod. Opt. 47 (2000) 155.

\bibitem{Nielsen}
A.~Gilchrist, N.~K.~Langford, M.~A.~Nielsen,
Phys. Rev. A 71 (2005) 062310.

\bibitem{boca}
M.~Boca, I.~Ghiu, P.~Marian, T.~A.~Marian,
Phys. Rev. A 79 (2009) 014302.

\bibitem{ghiu} I. Ghiu, G. Bj\"{o}rk, P. Marian, T. A. Marian, in: K. H\"{a}rk\"{o}nen, S. Maniscalco, J. Piilo,
K.-A. Suominen, O. Vainio (Eds.), Rep. Ser. Phys. Univ. Turku, Ser.
L 32, Uniprint, Turku, Finland, 2009, p. 115.

\bibitem{hiai} F. Hiai, M. Mosonyi, T. Ogawa, J. Math. Phys. 49 (2008) 032112.

\bibitem{unpublished} J.~S\"{o}derholm, \textit{et al.} unpublished.

\bibitem{sehat}
A.~Sehat, J.~S\"{o}derholm, G.~Bj\"{o}rk, P.~Espinoza, A.~B.~Klimov, L.~L.~S\'anchez-Soto,
Phys. Rev. A 71 (2005) 033818.

\bibitem{squeezing}
A.~Luis, N.~Korolkova, Phys. Rev. A 74 (2006) 043817.

\bibitem{carozzi}
T.~Carozzi, R.~Karlsson, J.~Bergman, Phys. Rev. E 61 (2000) 2024.

\bibitem{setala}
T.~Set\"{a}l\"{a}, K.~Lindfors, M.~Kaivola, J.~Tervo, A.~T.~Friberg,
Opt. Lett. 29 (2004) 2587.

\bibitem{luis 3D}
A.~Luis,
Phys. Rev. A 71 (2005) 063815.

\bibitem{karassiov}
V. P. Karassiov, Phys. Lett. A 190 (1994) 387.

\bibitem{rubin}
M. H. Rubin, D. N. Klyshko, Y. H. Shih, A.~V.~Sergienko, Phys. Rev.
A 50 (1994) 5122.

\bibitem{bushev}
P. A. Bushev, V. P. Karassiov, A.~V.~Masalov, A.~A.~Putilin, Opt.
Spectrosc. (USSR) 91 (2001) 526.

\bibitem{karassiov2}
V. P. Karassiov, S. P. Kulik, J. Exp. Theor. Phys. (USSR) 104 (2007)
30.

\end{thebibliography}
\end{document}